\documentclass[a4paper]{jpconf}
\usepackage{graphicx}
\usepackage{epsfig}
\usepackage{amsmath}
\usepackage{amssymb} 
\usepackage{subfig}
\usepackage{multirow}
\usepackage{float}
\usepackage{axodraw}
\newcommand {\pslash} {\slash \hspace{-0.26cm} P}
\newcommand {\kslash} {\slash \hspace{-0.22cm} k}

\newcommand {\hsp} {\hspace{0.45 cm}}

\def\date{\number\day\space de \ifcase\month\or
January,\or February,\or March,\or 
Appril,\or May,\or June,\or July,
\or Agust,\or septembre,\or Octubre,\or November,
\or December,\fi\space\number\year}

\begin{document}

\title{The pion electromagnetic structure with self-energy} 

\author{Clayton Santos Mello$^a$, J. P. B. C. de Melo$^b$ and T. Frederico$^a$}
\address{$^a$Instituto Tecnol\'ogico de
Aeron\'autica, DCTA, 12.228-900 S\~ao Jos\'e dos Campos, SP, Brazil.}
\address{$^b$Laborat\'orio de F\'\i sica Te\'orica e Computacional~-~LFTC 
Universidade Cruzeiro do Sul, 01506-000, S\~ao Paulo, SP, Brazil}
\ead{claytonsantosmello@yahoo.com.br}

\begin{abstract} 
We study the electromagnetic structure of the pion in terms of the 
quantum cromodynamic~(QCD) model on the Breit-frame. We calculated the observables, 
such as the electromagnetic form factor. 
The priori to have a calculation covariant need to get the valence term of 
the eletromagnetic form factor. We use the usual formalism in quantum field theory (QFT) and 
light-front quantum field theory (LFQFT) in order to test the  properties of form factor in nonperturbative QCD. 
In this particular case, the form factor can be obtained 
using the pion Light-Front (LF) wave function including self-energy from Lattice-QCD. Specifically, these calculations
 was performed in LF formalism. We consider a quark-antiquark vertex model having a 
 quark self-energy. Also we can use other models to  compare the pion electromagnetic form factor 
with different wave function and to observe the degree of agreement between them.
\end{abstract}

\section{Introduction}
 \hsp 
 The dynamics of the internal structure of hadrons affects their  observable properties, 
and the electromagnetic form factor of hadrons is an example of such an 
observable \cite{JVary22014,Bakker2011,PMaris2003,Bashir2003,
Rojas2013,Dudal2013,deMelo2004,deMelo2006,deMelo1997,
Maris1998}. The interaction of a virtual photon with a meson  probes its internal structure and dynamics 
through the meson form factor~\cite{CMello2015,deMelo2015,CMello2013,CMello2014}. The study of their form factors will thus allow us to 
extract information about the nonperturbative dynamics of its  
constituents~\cite{deMelo1997,CMello2014}. 
Specifically in this paper we 
treat only the pion meson internal structure. The theoretical prediction of 
electromagnetic form factor $F_\pi(q^2)$ at experimentally accessible $q^2$, 
below say $10 \ GeV^2$, is a nontrivial task since the complex 
nonperturbative physics of confinement \cite{JVary22014}, dynamical chiral symmetry 
breaking (DCSB), and bound state structure are highly 
dependent on the modelling of the strong coupling regime that is not reachable using perturbative-QCD.

 For our model we propose the inclusion of the self-energy of the quark at the 
 quark-antiquark vertex $\Gamma_\pi(k)$ in the chiral limit \cite{CRoberts2007}, 
 where the pion mass $m_\pi=0$. This account was used only to define this
 vertex. Since the calculation of the form factor was conducted in the 
 Breit-frame the vertex must be out of the chiral limit to
 satisfy the momentum conservation. The form factor was derived from the 
 Feynman triangular diagram at impulse approximation.
 So we have only one pion elastic form factor \cite{deMelo1997}.
 
 The valence wave function of the pion in our model comes from the 
 projection of the Bethe-Salpeter amplitude in the Light-Front~(LF), and integrated in
 $k^-$ (energy in LF). The four-vector momentum space in the LF, is 
 defined as,~$k^\mu$, where~$\mu=-,+,\perp$, energy, longitudinal and transversal momenta respectively.

 In the configurations space we have~$x^\mu$, where $\mu=-,+,\perp$, longitudinal position, 
 time and transversal direction respectively \cite{Harindranath96}.
 Here we use the valence wave function for the purpose of showing only the pion form factor of 
 valence whose is the square modulus of the wave function less the quark-photon 
 vertex $\Gamma^\mu(k,P)$. It will be
 constructed from Ward-Takahashi identity (WTI) \cite{Takahashi1957} in order to 
 ensure the momentum conservation for the electromagnetic current $J^\mu$.
 In order to minimize the zero-mode terms in LF \cite{TesedeMelo1998} we calculate only the 
 current component $J^+$. Since we have the electromagnetic current of 
 the pion we can get the electromagnetic form factor, as explained in the following sections.

\section{Pion and its constituent quarks}
 \hsp  
 Dynamical chiral symmetry breaking (DCSB) is one of the most important properties of low 
energy QCD \cite{zuber}, and its breaking pattern has profound impact on phenomenological quantities, e.g. the appearance of
 pseudoscalar Goldstone bosons \cite{Drell64} and the non-degeneracy of
chiral partners.
 The spontaneous breaking of chiral symmetry is a remarkable feature of QCD because 
it cannot be derived directly from the Lagrangian \cite{zuber} it is related to the nontrivial structure
 of the QCD vacuum, characterised by strong condensates of quarks and gluons \cite{CRoberts2007}. This is 
quite different from the explicit symmetry breaking, which is put in by hand through 
the finite quark masses, and appears in a similar way through the Higgs mechanism. 
There are two important consequences of the spontaneous breaking of chiral symmetry.
The first one is that the valence quarks acquire a dynamical or constituent mass 
through their interactions with the collective excitations of  the QCD vacuum that is 
much larger than the seed mass present in the Lagrangian. The second one is the appearance of a triplet 
of pseudoscalar mesons of low mass $(\pi^+, \pi^- , \pi^0)$  which represent
 the associated Goldstone bosons \cite{Halzen}.

 The prominent role played by the pion as the Goldstone boson of spontaneously 
broken chiral symmetry has its impact on the low-energy structure of hadrons through
pion cloud effects in the quark propagation \cite{Lei2013}. In full QCD there are hadron contributions
to the fully dressed quark-gluon vertex. These effects are generated by the inclusion of
dynamical sea quarks in the quark-gluon interaction, and are therefore only present in
the unquenched case. It is the aim of this paper to introduce these pion cloud effects
into the quark propagator through quark mass function, and then all the way up into the 
meson Bethe-Salpeter amplitude (BSA) and the pion electromagnetic form factor. 

\section{Self-energy into pion-quark-antiquark vertex} 
\hsp
 Due to the inclusion of quark self-energy at vertex we have a  
 pseudoscalar pion vertex it is given by \cite{CRoberts2007}:
\begin{eqnarray}
 \Gamma_\pi(k)=i \gamma^5 M(k) \ ; \ \text{where} 
 \ M(k)=m_0- \dfrac{m^3}{k^2 - \lambda^2-i \epsilon} \ ,
\end{eqnarray}
 where $\gamma^5$ is the Dirac matrix, $k$ is the relative momentum between the 
 constituent quarks of the pion. $M(k)$ is the quark mass function 
 that has been obtained from Schwinger-Dyson equation (SDE) solutions~\cite{Rojas2013,Beane2011} who was able to fit the
 Lattice-QCD calculations~\cite{Montvay1994,Sterman1997,Muller1999}.
 The $m_0=0.014 \ GeV$ is the current quark mass and $m^3=0.189 \ GeV^3$ 
 and $\lambda^2=0.639 \ GeV^2$ are the Lattice-QCD parameters \cite{Dudal2013}. 
The quark propagator also contains the quark mass function due to the 
presence of the self-energy in the legs into pion-quark-antiquark vertex. 
So we have to the quark propagator in the LF coordinates:

\begin{eqnarray}
  S(k) &=&  \dfrac{\kslash+M(k)}{k^2-M^2(k) +i \epsilon} 
 = \dfrac{\kslash_{\text{on}}+M(k)}{k^2-M^2(k) +i \epsilon}+ \dfrac{\gamma^+}{2k^+} \ , 
\end{eqnarray} 
 In the expression above, 
 the~$\kslash_{\text{on}}$ subscript indicates the quark is on-shell. 
 
And that has been obtained by the separation 
 $\frac{\gamma^+}{2k^+}$ instant term in LF. The $k^-_{\text{on}}= \frac{k^2_\perp+M^2(k)}{k^+}$, it is on-shell energy.
 The four-vector in LF is defined for the usual coordinates as: $k^\pm=k^0 \pm k^3$ and $k_\perp=(k^1,k^2)$ the same for the Dirac matrices. The consequence of this is that the dot product $k^2=k^+k^--k^2_\perp$ \cite{Harindranath96}. 
The increase $i \epsilon$ allows us to delocate the poles that contribute to integration on $k^-$ via Cauchy's theorem \cite{TesedeMelo1998}.
 This calculation makes the relative time between quarks to be eliminated \cite{deMelo1997}.
 
\subsection{Quark-photon vertex} 

\hsp
For the Dirac structure of the electromagnetic current, we make use of Ward-Takahashi identity \cite{Takahashi1957}. 
So we can extract the quark-photon vertex as follows: 
\begin{eqnarray} 
 q_\mu \Gamma^\mu(k;P,P') &=& S^{-1}(P'-k) - S^{-1}(P-k) \nonumber \\
\Gamma^\mu(k;P,P') &=& \gamma^\mu + \Lambda^\mu (k;P,P') \ ,
\end{eqnarray}
where the correlation function, $\Lambda^\mu (k,P)$, is given by:
\begin{eqnarray}
\Lambda^\mu (k;P,P') =  \dfrac{ m^3 \left(2 k-P'- P \right)^\mu
}{ \left[ (P'-k)^2- \lambda^2+i \epsilon \right]
\left[ (P-k)^2- \lambda^2+i \epsilon \right]} \ . 
\end{eqnarray}

 From the Feynman triangular diagram, we obtain the pion electromagnetic current $J^\mu$:
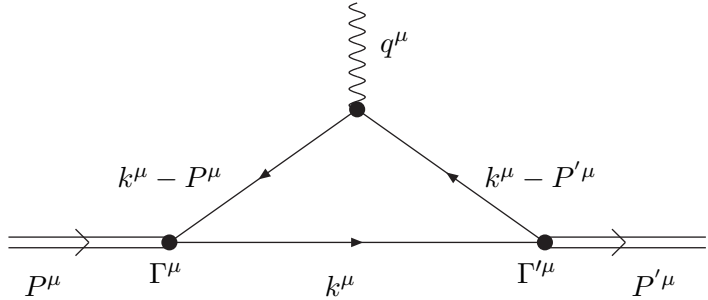
\begin{figure}[h]
\begin{center}
\vskip 0.5 cm
\begin{picture}(330,130)(0,0)
\Line(0,52)(60,52)
\Line(25,55)(30,50)
\Line(25,45)(30,50)
\Line(0,48)(60,48)
\Vertex(60,50){3.0}
\put(65,35){\makebox(0,0)[br]{$ \Gamma^\mu $}}
\ArrowLine(130,100)(60,50)
\ArrowLine(60,50)(200,50)
\Vertex(200,50){3.0}
\put(205,35){\makebox(0,0)[br]{$ \Gamma'^\mu $}}
\ArrowLine(200,50)(130,100)
\Line(200,48)(260,48)
\Line(225,55)(230,50)
\Line(225,45)(230,50)
\Line(200,52)(260,52)
\Photon(130,100)(130,140){3}{6.5}  
\put(150,120){\makebox(0,0)[br]{$ q^\mu $}} 
\Vertex(130,100){3.0}
 \put(80,70){\makebox(0,0)[br]{$ k^\mu-P^\mu $}}
 \put(20,30){\makebox(0,0)[br]{$ P^\mu $}}
 \put(130,30){\makebox(0,0)[br]{$ k^\mu $}}
 \put(220,70){\makebox(0,0)[br]{$ k^\mu -P^{' \mu} $}}
 \put(250,30){\makebox(0,0)[br]{$ P^{' \mu} $}}
\end{picture}
\vspace{-1.cm}
   \caption{ Triangular diagram in impulse  approximation.}
\label{fig9} 
 \end{center}
\end{figure}
%
\begin{eqnarray}
  J^\mu &=& i \dfrac{N^2N_c}{f_{\pi}^2} \int \dfrac{d^4 k}{(2 \pi)^4}  
  Tr\left[ S(k) \Gamma_\pi(P'-k) S(P'-k) \Gamma^\mu(k;P,P') S(P-k) 
  \Gamma_\pi(P-k) \right] \nonumber \\
 J^\mu &=& i \dfrac{N^2N_c}{f_{\pi}^2} \int \dfrac{d^4 k}{(2 \pi)^4}  
 \dfrac{Tr \left[\mathcal{O}^\mu(k;P,P') \right]}
 {\left[k^2- M^2(k)+i \epsilon \right]}  M(P'-k;m_0=0) M(P-k;m_0=0) \nonumber \\
   && \dfrac{1}{\left[(P-k)^2- M^2(P-k)+i \epsilon \right]
   \left[(P'-k)^2- M^2(P'-k)+i \epsilon \right]} \ .
\end{eqnarray} 
 
 Now we might writte the Dirac structure of this way:
\begin{eqnarray}
 \mathcal{O}^\mu(k;P,P') &=&  \left(\kslash+M(k)\right) \gamma^5 \left(\pslash '-\kslash+M(P'-k)\right) \Gamma^\mu(k;P,P')  \nonumber \\
 && \left(\pslash-\kslash+M(P-k)\right) \gamma^5 \ , 
\end{eqnarray}
where $q=P'-P$, and using the Breit frame that is the  transfer 
momentum in x-direction $q = (0,q^x,0,0)$. To the total initial momentum 
of the pion it is $P=(P^0,-q^x/2,0,0)$ and total final momentum is $P'=(P^0,q^x/2,0,0)$. We
 also made use of Drell-Yan condition, $q^+=0$, when there no  transfer momentum on longitudinal 
 direction \cite{TesedeMelo1998,Silva2012,Yabuzaki2015}. The Dirac trace structure, is writing as,  
\begin{eqnarray}
   \mathcal{P}^+_n(k,P,P')&=&  -4m^6 \left[k^+ \left(a_n-b_n-c_n \right) +\left(b_nP^++c_nP'^+\right) \right]   \nonumber \\ 
  &+& 4m_0m^3\left[\left(c_nP'^++b_nP^+ \right)a_n+\left(-2k^++P^++P'^+ \right)b_nc_n \right] \nonumber \\
  &+& m_0 \left[4 \left(k^2_x+m_0^2 \right)+2 \left(P'^-P^++P^-P'^+ \right)+q^2 \right] \times \nonumber \\
  & \times & a_n(2k^+-P^+-P'^+)m^3 \nonumber \\
  &+& a_nb_nc_n \left \{\left[-4 \left[k^-_n(k^+-P^+)(k^+-P'^+) \right. \right. \right. \nonumber \\
  &+& \left. k_x^2(P^++P'^+)\right]-2k_x (P'^+-P^+)q \nonumber \\
  &+& \left. \left. k^+(4k_x^2+q^2) \right]+4m_0^2\left[k^+-P^+-P'^+ \right]  \right \} \nonumber \\
  &+& \left \{ \left[-2k^+P'^-a_nc_n+2k_xq(c_n-b_n)a_n-2k^+P^-a_nb_n \right. \right. \nonumber \\
  &-&  4 k_x^2(a_n(b_n+c_n)-b_nc_n)+2k^+b_nc_n(P'^-+P^-) \nonumber \\
  &-& 4m_0^2(a_n(b_n+c_n)+b_nc_n) \nonumber \\
  &+& \left. b_nc_n(-2P'^-P^+-2P^-P'^+-q^2)  \right]m^3 \nonumber \\
  &+& 2k^-_n \left[\left(-P'^+a_nc_n-P^+a_nb_n \right)+2k^+(a_n(b_n+c_n)-b_nc_n) \right. \nonumber \\
  &+&   \left. (P^++P'^+)b_nc_n \right]m^3-4m^6 \left[m^3-m_0 \left(a_n+b_n+c_n \right) \right]  \nonumber \\
  &-& \left. 4m_0 k_n^-k^+a_nb_nc_n \right \}  \frac{(2k^+-P^+-P'^+)m^3}{b_nc_n} \nonumber \\
  Tr[\mathcal{O}^+_n(k;P,P')]&=& \dfrac{\mathcal{P}^+_n(k;P,P')}{a_nb_nc_n} \ .
\end{eqnarray}
 Thus we have the electromagnetic current this way:
\begin{eqnarray}
 J^+ &=& i \dfrac{N^2N_c}{f_{\pi}^2} 
 \int \dfrac{d^4 k}{(2 \pi)^4} 
 \dfrac{m^6Tr \left[\mathcal{O}^+(k;P,P') \right]a^2_nb^2_nc^2_n}{\left(a^2_n(k^2+i \epsilon) 
 -(m_0a_n- m^3)^2) \right)b_nc_n}  \nonumber \\
   & & \dfrac{1}{\left(b^2_n((P-k)^2+i \epsilon)-(m_0b_n- m^3)^2) \right)} \nonumber \\
   & &  \dfrac{1}{\left(c^2_n((P'-k)^2+i \epsilon)-(m_0c_n- m^3)^2) \right)}  \nonumber \\ 
& = & i \dfrac{N^2N_c}{f_{\pi}^2} \int \dfrac{d^4 k}{(2 \pi)^4} \dfrac{m^6 \mathcal{P}^+(k;P,P')
  (y- \lambda^2+i \epsilon) }{(y-y_1)(y-y_2)(y-y_3)}  \nonumber \\
   && \dfrac{1}{(z-z_1)(z-z_2)(z-z_3)(w-w_1)(w-w_2)(w-w_3)} \ . \   
\end{eqnarray}  
 Now we have written the quark mass function below, in order to obtain an expression for the pion electromagnetic form factor:
\begin{eqnarray}
  M(k)&=&m_0-\dfrac{m^3}{a_n} \ , \ a_n=k^+ \left(k^-_n- \dfrac{f_a-i \epsilon}{k^+} \right) \ , \nonumber \\
  M(P-k)&=&m_0-\dfrac{m^3}{b_n} \ , \
  b_n= (P^+-k^+) \left(P^--k^-_n- \dfrac{f_b-i \epsilon}{(P^+-k^+)}\right) \ , \ \nonumber \\  
    M(P'-k)&=& m_0-\dfrac{m^3}{c_n} \ , \
   c_n = (P'^+-k^+) \left(P'^--k^-_n- \dfrac{f_c-i \epsilon}{(P'^+-k^+)} \right) \ , 
\end{eqnarray}
where:
\begin{eqnarray}
  f_1 &=& k_\perp^2+Re[y_1] \ , \
  f_2 = k_\perp^2+Re[y_2] \ , \
  f_3 = k_\perp^2+Re[y_3] \ , \
  f_a = k^2_\perp+ \lambda^2 \nonumber \\
  f_4 &=& (P-k)^2_\perp+Re[z_1] \ , \
  f_5 = (P'-k)^2_\perp+Re[z_2] \ , \
  f_6 = (P-k)^2_\perp+Re[z_3] \nonumber \\
  f_7 &=& (P'-k)^2_\perp+Re[w_1] \ , \
  f_8 = (P-k)^2_\perp+Re[w_2] \ , \nonumber \\
  f_9 &=& (P'-k)^2_\perp+Re[w_3] \ , \   
  f_b = (P-k)^2_\perp+ \lambda^2 \ , \
  f_c = (P'-k)^2_\perp+ \lambda^2 \ .
\end{eqnarray}
where $z_1=w_1=y_1$, $z_2=w_2=y_2$ and $z_3=w_3=y_3$ are the 
denominator roots of the electromagnetic current and they are derived of propagators when  including the self-energy in legs at pion vertex. 
We identify the poles in $k^-$ when we perform a change of  
 variable $k^2=y$, $(P-k)^2=z$ and $(P'-k)^2=w$, and 
 the roots are: 
 \begin{eqnarray}
  e &=& -6 m^3m_0-2 \lambda^2 m_0^2+m_0^4   \ , \nonumber \\
  f &=& -18 \lambda^2 m^3 m_0 -6 \lambda^4 m_0^2+18 m^3 m_0^3
  +6 \lambda^2 m_0^4 - 2 m_0^6   \ , \nonumber \\
  g &=& 36 \lambda^2 m^9 m_0+8 \lambda^4 m^6 m_0^2-4 m^9 m_0^3
  -4 \lambda^2 m^6 m_0^4   \ , \nonumber \\
  d &=&\sqrt[3]{2 \lambda^6-27m^6+3 \sqrt{3} \sqrt{-4 \lambda^6m^6+27m^{12}+g}+f}   \ , \nonumber \\
  y_1&=&\dfrac{2}{3} \lambda^2+ \dfrac{m_0^2}{3} - \dfrac{\sqrt[3]{2}}{3d} \left( \lambda^4+e \right)
  - \dfrac{d}{3 \sqrt[3]{2}} -i \epsilon   \ , \nonumber \\
  y_2&=&\dfrac{2}{3} \lambda^2+ \dfrac{m_0^2}{3}+ 
  \dfrac{(1+i \sqrt{3})}{3 \sqrt[3]{4}d } \left(\lambda^4+e \right) + 
  \dfrac{(1-i \sqrt{3})d}{6 \sqrt[3]{2}}-i \epsilon   \ , \nonumber \\
   y_3&=&\dfrac{2}{3} \lambda^2+ \dfrac{m_0^2}{3}+ \dfrac{(1-i \sqrt{3})}{3 \sqrt[3]{4}d } 
   \left(\lambda^4+e \right) + \dfrac{(1+i \sqrt{3})d}{6 \sqrt[3]{2}}-i \epsilon   \ .  
\end{eqnarray}
 
 The electromagnetic current,$J^+$, is also, writting like, 
\begin{eqnarray}
 J^+ &=& i \dfrac{N^2N_c}{f_{\pi}^2} \int \dfrac{d^2 k_\perp 
 dk^+dk^-}{2(2 \pi)^4} \dfrac{m^6 \mathcal{P}^+_n(k;P,P') a_n}
 {\mathcal{D}_1\mathcal{D}_2\mathcal{D}_3\mathcal{D}_4
\mathcal{D}_5\mathcal{D}_6\mathcal{D}_7\mathcal{D}_8\mathcal{D}_9}~. 
\end{eqnarray}
 here, are, 
 \begin{eqnarray*}
 \mathcal{D}_1 &=& k^+\left(k^-_n-\dfrac{f_1 -i \epsilon}{k^+} \right);~
  \mathcal{D}_2 = k^+\left(k^-_n-\dfrac{f_2 -i \epsilon}{k^+} \right);
  \nonumber \\ 
   \mathcal{D}_3 & =&  k^+\left(k^-_n-\dfrac{f_3 -i \epsilon}{k^+} \right);~
\mathcal{D}_4 = \left(P^+ -k^+ \right) \left(P^- - k^-_n - \dfrac{f_4 -i \epsilon}{P^+-k^+} \right); 
\nonumber \\  
\mathcal{D}_5 &=& \left(P'^+ -k^+ \right) \left(P'^- - k^-_n - \dfrac{f_5 -i \epsilon}{P'^+-k^+} \right);~  
\mathcal{D}_6 =  \left(P^+ -k^+ \right) \left(P^- - k^-_n - \dfrac{f_6 -i \epsilon}{P^+-k^+} \right);~ 
\nonumber \\  
\mathcal{D}_7 &=& \left(P'^+ -k^+ \right) \left(P'^- - k^-_n - \dfrac{f_7 -i \epsilon}{P'^+-k^+} \right); 
\mathcal{D}_8 = \left(P^+ -k^+ \right) \left(P^- - k^-_n - \dfrac{f_8 -i \epsilon}{P^+-k^+} \right);~
\nonumber \\  
 \mathcal{D}_9 &=& \left(P'^+ -k^+ \right) \left(P'^- - k^-_n - \dfrac{f_9 -i \epsilon}{P'^+-k^+} \right)~. 
\label{jp}
\end{eqnarray*}
We can identify nine propagators in electromagnetic current equation, $J^+$,~with the following poles: 
\begin{eqnarray}
 k^-_1 & = &  \dfrac{f_1}{k^+}-\dfrac{i \epsilon}{k^+},~
 k^-_2  =  \dfrac{f_2}{k^+}-\dfrac{i \epsilon}{k^+},   k^-_3  =   \dfrac{f_3}{k^+}-\dfrac{i \epsilon}{k^+},
 \nonumber  \\
 k^-_4  & = &    P^-- \dfrac{f_4}{P^+-k^+}+\dfrac{i \epsilon}{P^+-k^+},~ 
k^-_5    =    P'^-- \dfrac{f_5}{P'^+-k^+}+\dfrac{i \epsilon}{P'^+-k^+},~ 
\nonumber \\
  k^-_6  & = & P^-- \dfrac{f_6}{P^+-k^+}+\dfrac{i \epsilon}{P^+-k^+}, 
    k^-_7  =  P'^-- \dfrac{f_7}{P'^+-k^+}+\dfrac{i \epsilon}{P'^+-k^+},~
  \nonumber \\
 k^-_8 & = &  P^-- \dfrac{f_8}{P^+-k^+}+\dfrac{i \epsilon}{P^+-k^+},
 k^-_9  =  P'^-- \dfrac{f_9}{P'^+-k^+}+\dfrac{i \epsilon}{P'^+-k^+}.
\end{eqnarray}

\begin{figure}[htb]
\vskip -0.5 cm
\begin{center}
\centerline{
\begin{picture}(330,130)(0,0)
\put(230,115){\makebox(0,0)[br]{$(i) \ \ \ \ \ 0<k^+<P^+ $}}
\put(162,100){\makebox(0,0)[br]{$Im [k^-]$}}
\put(320,35){\makebox(0,0)[br]{$Re [k^-]$}}
\LongArrow(100,55)(290,55)
\LongArrow(120,30)(120,110)
\Vertex(146,43){2}
\put(155,28){\makebox(0,0)[br]{$k^-_{1}$}}
\Vertex(176,43){2}
\put(185,28){\makebox(0,0)[br]{$k^-_{2}$}}
\Vertex(206,43){2}
\put(215,28){\makebox(0,0)[br]{$k^-_{3}$}}
\Vertex(128,78){2}
\put(137,60){\makebox(0,0)[br]{$k^-_{4}$}}
\Vertex(149,78){2}
\put(158,60){\makebox(0,0)[br]{$k^-_{5}$}}
\Vertex(170,78){2}
\put(179,60){\makebox(0,0)[br]{$k^-_{6}$}}
\Vertex(191,78){2}
\put(199,60){\makebox(0,0)[br]{$k^-_{7}$}}
\Vertex(211,78){2}
\put(220,60){\makebox(0,0)[br]{$k^-_{8}$}}
\Vertex(236,78){2}
\put(245,60){\makebox(0,0)[br]{$k^-_{9}$}}
\end{picture}
}
\end{center}
\vskip -1.8 cm
\caption{Poles position in Argand-Gauss plane for valence term of the electromagnetic current.} 
\label{fig2}
\end{figure}
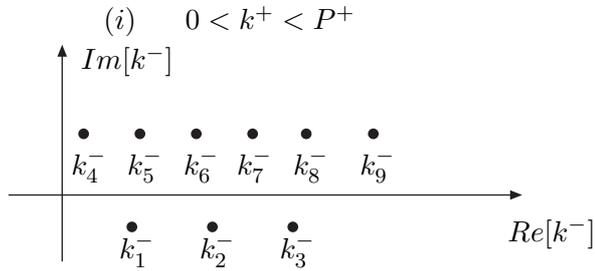

 To valence range of the electromagnetic current, we verified the poles contribution, Fig. (\ref{fig2}).

We have the valence contribuition term for the electromagnetic current, $J^{+V}$, 
which correspond the interval integration,~$0<k^+<P^+$,~in the light-front energy, $k^-$,
(see the Fig.(\ref{fig2}) above),~is given below, 
\begin{eqnarray}
 J^{+V} &=& \dfrac{N^2N_c}{f_{\pi}^2} \sum_{n=1}^3 \int 
 \dfrac{d^2 k_\perp dk^+}{2(2 \pi)^3} \dfrac{m^6 \mathcal{P}^+_n \mathcal{D}_n}
{k^+ \mathcal{D}_1\mathcal{D}_2\mathcal{D}_3\mathcal{D}_4
\mathcal{D}_5\mathcal{D}_6\mathcal{D}_7\mathcal{D}_8\mathcal{D}_9}~,
\end{eqnarray}

 
From the electromagnetic current, we can obtain the pion space-like electromagnetic form factor, with the 
expression below:
\begin{eqnarray}
  \left<P'^+ \right|J^+ \left|P^+ \right>=
  e(P'^++P^+) F_\pi(q^2) \ .
\end{eqnarray}
where $e$ is the elementary charge, and $F_\pi(q^2)$ is the electromagnetic form factor; 
the constant normalization $N$ is obtained from the condition
of charge,~$F_\pi(q^2=0)=1$,~\cite{deMelo2004,deMelo2006,TesedeMelo1998,Pacheco99,Pacheco2002}. 

\section{Numerical results}
\begin{figure}
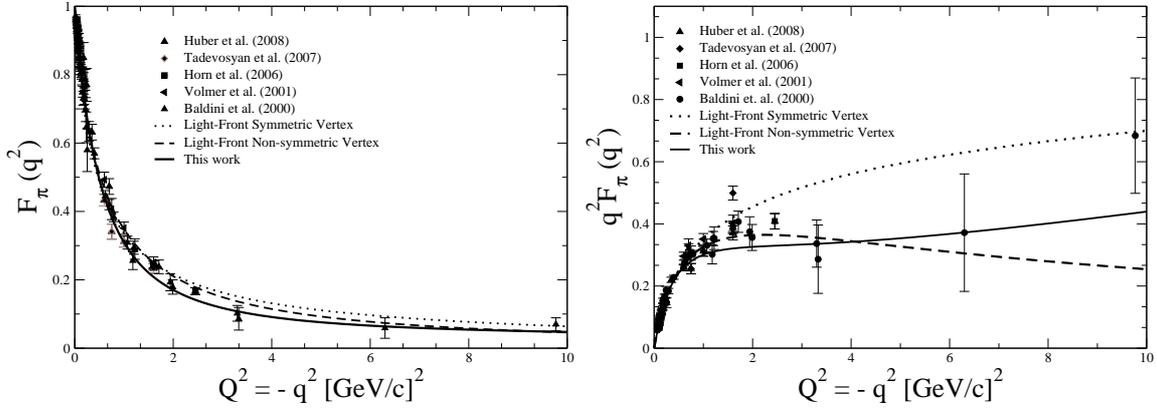

\begin{center}
\includegraphics[scale=0.3]{ffnp.eps}  
\includegraphics[scale=0.3]{q2ffnp.eps}
\end{center} 
\caption{Pion electromagnetic form factor in terms of momentum transfer,~$q^2$, 
with the light-front symmetric model,~point line~\cite{Pacheco2002}, 
dashed line, light-front non-symmetric vertex~\cite{Pacheco99}, solid line, 
present work, with the parameters from Lattice-QCD \cite{Dudal2013} and the experimental~\cite{Baldini2000}. 
 In this work, we use  the following values parameters from the lattice-QCD, 
 i.e, $\lambda^2=0.639~GeV^2,~m_0=0.014~GeV$ and $m^3=0.189~GeV^3$. 
 }
 \label{ffnp}
\end{figure}

\hsp
 For analysis of our model we present the results we obtained for the electromagnetic form factor of the pion.
 In the Fig. (\ref{ffnp}) on the left we can see the form factor as a function of the square transfer momentum. And
 on right we find the same function multiplied by $q^2$.
 Also we compare our results with two other models, they 
 are Light-front symetric vertex \cite{Pacheco2002} and non-symetric vertex \cite{Pacheco99}. Since the
 technique used in our calculations is also based on these models have drawn analytical valence wave function
 to separate the instant terms in LF.
 We compared our model for the form factor with the experimental data according to the references \cite{Volmer2001,Horn2006,Tadevosyan2007,Huber2008}.
 These data are for describing the structure of the pion at low energies. In this case $q^2$ for values lower than $10 \ [GeV/c]^2$.

\section{Conclusion}
\hsp
 In this paper, we show how to describe the electromagnetic structure of a particle pseudoscalar, pion.
 Through a constituent quark model, we obtained some of the pion observable, but using the formalism in light-front
 we see across terms that are not invariant under Lorentz transformations. In our case we do not find non-valence terms
 to form factor when it is in Breit-frame and on Drell-Yan condition.

 We study the electromagnetic current in respect to the valence term which we can extract the wave function of two
 fermions system, the quark-antiquark pair that composes the pion structure. From the wave function at the center of mass 
 of the system, technique that has identified valence wave function in our model, we get the observable of 
 pion, like the electromagnetic form factor. The electromagnetic current is decribed by constituent quarks 
 model with a self-energy at 
 pion vertex as expected. The form factor decreases with increasing of the transfer momentum. In relation to the
 form factor multiplied by square transfer momentum, it increases with increasing of $q^2$. The quark mass function
 describes the system in which particles dynamically gain mass, according to the constituent quark model. From the
 contribution of the photon we can study the internal  structure of the pion as pion electromagnetic structure.

\end{document}